\documentclass[
    ,final            
  ]
  {aipproc}

\layoutstyle{6x9}

\usepackage{latexsym,graphicx,rotating,amsmath, epsfig}
\usepackage{natbib}
\def\sol{\odot}

\begin{document}

  \title{Strong Dynamo Action in Rapidly Rotating Suns}



  \author{Benjamin P.\ Brown}{address={JILA and Dept.\ of Astrophysical \& Planetary Sciences, University of Colorado, Boulder, CO 80309-0440}}
  \author{Matthew K. Browning}{address={Dept.\ of Astronomy, University of California, Berkeley, CA 94720-3411}}
  \author{Allan Sacha Brun}{address={DSM/DAPNIA/SAp, CEA Saclay, Gif sur Yvette, 91191 Cedex, France}}
  \author{Mark S. Miesch}{address={High Altitude Observatory, NCAR, Boulder, CO 80307-3000}}
  \author{Nicholas J. Nelson}{address={JILA and Dept.\ of Astrophysical \& Planetary Sciences, University of Colorado, Boulder, CO 80309-0440}}
  \author{Juri Toomre}{address={JILA and Dept.\ of Astrophysical \& Planetary Sciences, University of Colorado, Boulder, CO 80309-0440}}

\begin{abstract}
    Stellar dynamos are driven by complex couplings between rotation
    and turbulent convection, which drive global-scale flows and build
    and rebuild stellar magnetic fields.  When stars like our sun are
    young, they rotate much more rapidly than the current solar rate.
    Observations generally indicate that more rapid rotation is
    correlated with stronger magnetic activity and perhaps more effective
    dynamo action.  Here we examine the effects of more rapid rotation
    on dynamo action in a star like our sun.  We find that vigorous
    dynamo action is realized, with magnetic field generated
    throughout the bulk of the convection zone.  These simulations do
    not possess a penetrative tachocline of shear where global-scale
    fields are thought to be organized in our sun, but despite this we
    find strikingly ordered fields, much like sea-snakes of toroidal
    field,  which are organized on global scales.
    We believe this to be a novel finding.

\end{abstract}

\maketitle

\subsection{1. Coupling of convection, rotation and magnetism}
Rotation and convection are key components of stellar dynamo action.
It is their complex coupling which must lead to the global-scale
fields observed in our sun and other solar-like stars.
When stars like our sun are young they rotate much more rapidly than
the current solar rate.  Observations generally indicate that more
rapid rotation is correlated with stronger magnetic activity, which
may indicate a stronger stellar dynamo.  Here we explore the effects
of more rapid rotation on convection and dynamo action in a more
rapidly rotating solar-like star.

In the sun, global-scale dynamo action is thought to arise from the
coupling of convection and rotation and the resulting global-scale
flows of differential rotation and meridional circulation.  As
revealed by helioseismology, the solar
differential rotation profile observed at the surface prints
throughout the bulk of the convection zone, with two prominent regions of radial
shear.  The near-surface
shear layer exists in the outer 5\% of the sun, whereas the tachocline lies
between the radiative interior, which is in nearly solid body
rotation, and the convective envelope above \citep{Thompson_et_al_2003}.  In the
interface dynamo model \citep[e.g.,][]{Charbonneau_2005}, magnetic
fields generated in the bulk of the convection zone are pumped into
the stably stratified tachocline where the strong radial shear builds
and organizes the global-scale fields that eventually erupt at the
solar surface.  The differential rotation plays an important role in
the production of the global-scale magnetic fields while the meridional
circulations may be important for returning flux to the base of the
convection zone, enabling cycles of magnetic activity.  In
observations of solar-like stars the differential rotation appears to
grow stronger at more rapid rotation rates
\citep[e.g.,][]{Donahue_et_al_1996}.  In our initial exploration of
dynamo action in more rapidly rotating suns, we have undertaken simulations
which self-consistently establish differential rotation through the
coupling of convection and rotation and then explore the resulting
dynamo action in the bulk of the convection zone.

\subsection{2. Global simulations of stellar convection}
\label{sec:ASH}
To capture the essential couplings between convection, rotation and
magnetism, we must employ a global model which 
simultaneously captures the spherical shell geometry and admits the
possibility of zonal jets, large eddy vortices and convective plumes
which may span the depth of the convection zone, as well as
global-scale magnetic structures.  Stellar convection
zones are intensely turbulent and molecular values of viscous, thermal
and magnetic diffusivity in stars are estimated to be very small.  As a
consequence, numerical simulations cannot hope to resolve all scales
of motion present in real stellar convection and a compromise must be
struck between faithfully capturing the important dynamics within
small regions and capturing the connectivity and geometry of the
global scales.  

Our tool for exploring stellar dynamos is the anelastic spherical
harmonic (ASH) code, which is described in detail in
\cite{Clune_et_al_1999} and with magnetism in \cite{Brun_et_al_2004}.  
ASH is a mature simulation code, designed to run on massively parallel
architectures, which solves the three-dimensional MHD equations of
motion under the anelastic approximation.  The treatment of velocities
and magnetic fields is fully non-linear, 
but under the anelastic approximation the thermodynamic
variables are linearized about their spherically symmetric and
evolving mean state with density $\bar{\rho}$, pressure $\bar{P}$,
temperature $\bar{T}$ and specific entropy $\bar{S}$ all varying with
radius.  The anelastic approximation captures the effects of density
stratification but filters out sound waves and the fast
magneto-acoustic waves which would severely limit the time step.
These acoustic waves are largely decoupled from the decidedly subsonic
convective motions in the interior.

With present and foreseeable computational resources, global-scale
codes cannot resolve all scales of motion present in real stellar
convection zones.  With ASH, we explicitly resolve the largest
scales of motion and model the transport properties of scales below
our resolution.  We are thus performing a large eddy
simulation (LES) with subgrid-scale (SGS) modelling. The current SGS
model treats these scales with effective eddy diffusivities $\nu$, $\kappa$, and $\eta$,
representing the transport of momentum, heat and magnetic field by
the unresolved motions.  For simplicity these diffusivities
are taken as functions of radius alone and in these simulations
are proportional to $\bar{\rho}^{-1/2}$.  

As we are primarily interested in the coupling of rotation, convection
and global-scale flows in the bulk of the convection zone, we avoid
the H and He ionization regions near the stellar surface as well as
the tachocline of shear at the base of the convection zone.  Our
simulation extends from $0.72 R_\sol$ to $0.96 R_\sol$, with an
overall density contrast of 40 across the domain.  Solar values are
taken for heat flux, mass and radius, and a perfect gas is assumed.
The reference state of our thermodynamic variables is derived from a
one-dimensional solar structure model \citep{Brun_et_al_2002} and is
continually updated with the spherically-symmetric components of the
thermodynamic fluctuations as the simulations proceed.  The
simulations reported here are of a star rotating at three times the
current solar rate and are derived from hydrodynamic case~G3 as
reported in \cite{Brown_et_al_2007}, and these dynamo simulations are
denoted as case~D3. An initial weak dipole field was introduced and
over a 1000 day period the simulation amplified the total magnetic
energy by several orders of magnitude.  The simulation then ran an
additional 4000 days in this equilibrated state to explore the
long-term dynamo behaviour.  The mid-convection zone value of $\nu$ is
$1.32 \times 10^{12}~\mathrm{cm}^2/\mathrm{s}$ and the Prandtl number
$Pr=\nu/\kappa$ is 0.25 while the magnetic Prandtl number $Pm =
\nu/\eta$ is 0.5.  The ohmic diffusion time at mid-convection zone
$\tau_\eta = D^2/\eta$ is approximately 1200 days, with $D$ the depth
of the convection zone.  Our upper boundary is stress free for
velocities and potential field for magnetism.  The lower boundary is
stress free and a perfect electric conductor.  The spatial resolution
in these simulations involves using spherical harmonics up to degree
170 or 340 with typically 96 to 192 radial grid points, and the study
of the temporal evolution requires close to five million timesteps.

\begin{figure*}[t]
  \includegraphics{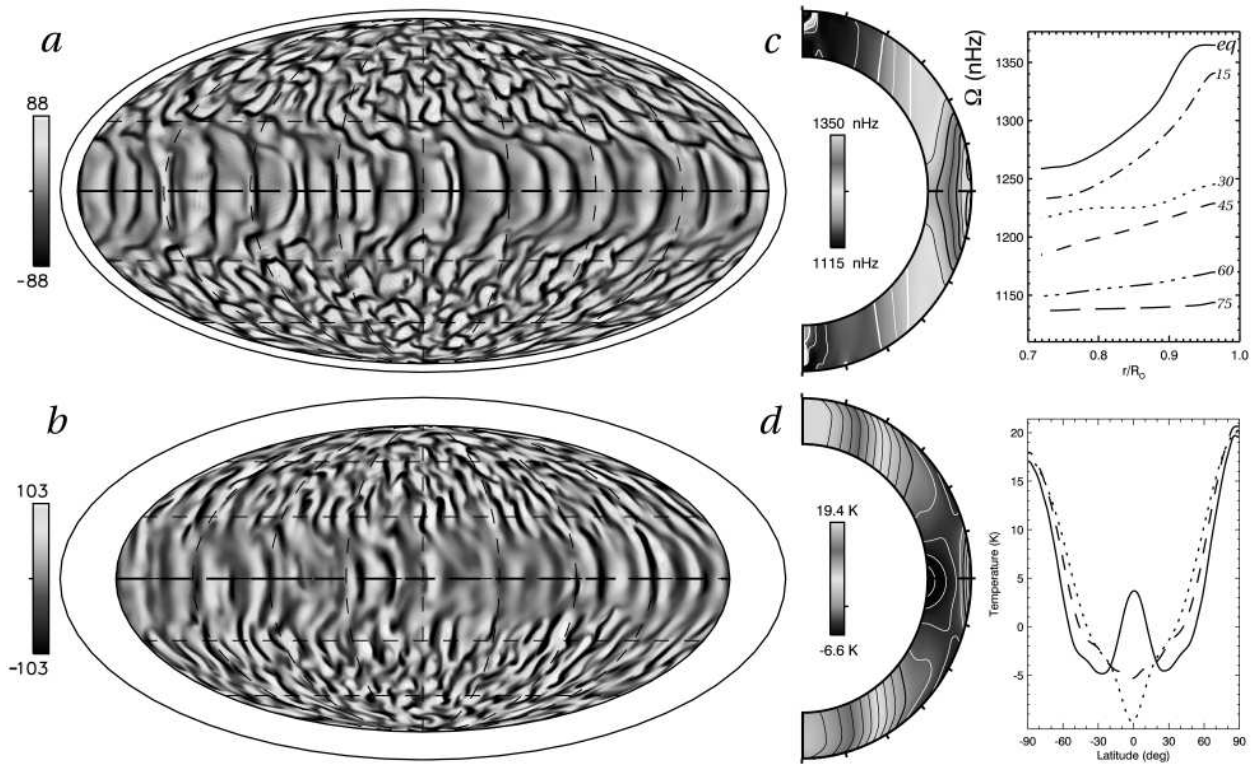}
  \caption{Snapshots of radial velocity in global Mollweide view 
    at the same instant in time $(a)$ near the stellar surface
    ($r=0.95 R_\sol$) and $(b)$ at mid-convection zone ($r=0.85
    R_\sol$) for case~D3.  Upflows are denoted by light tones while downflows are
    dark, with scale in m/s.  Strong plumes penetrate to the
    deepest layers, while weaker plumes are double-celled in
    radius. $(c)$ Azimuthal average of angular velocity $\Omega$ with
    radius and latitude.  This has been
    further averaged in time over a 240 day period. The equator is fast
    and the poles are slow.  Plotted at right are radial cuts of
    $\Omega$ at selected latitudes as indicated. 
    $(d)$ Azimuthal and time average of temperature fluctuations $T'$ about the
    spherically symmetric mean $\bar{T}$.  The poles are warm and the
    equator largely cool.  Plotted at right are three cuts in
    latitude near top ($r=0.96 R_\sol$, solid, with
    $\bar{T}=2.7\times10^{5}~\mathrm{K}$)
    , middle ($r=0.85 R_\sol$,
    long dash, with $\bar{T}=1.1\times10^{6}~\mathrm{K}$) and bottom of the convection zone ($r=0.72 R_\sol$,
    short dash, with $\bar{T}=2.3\times10^{6}~\mathrm{K}$).  Although
    distinctive, the latitudinal temperature contrast of order 20K is
    but a very small fraction of the spherically-symmetric mean temperature.
    \label{fig:vr_dr_and_T}}
\end{figure*}
\subsection{3. Structure of convection}
The structure of convective patterns in our more rapidly rotating sun
is illustrated in Figure~\ref{fig:vr_dr_and_T}.  The radial velocity
is shown near the top of the domain and at mid-convection zone in
Mollweide projection, with poles at top and bottom and the entire
equatorial region in the middle.  
Asymmetries in the convection arise from the density stratification,
resulting in narrow, fast downflows surrounded by broad, weaker upflows.
Convection in the equatorial region is
dominated by large cells aligned with the rotation axis.  These
convective structures are largely double-celled in radius, owing to
the strong radial shear in differential rotation, but strong
intermittent downflows span the domain.  At high latitudes the convection
is more isotropic and cyclonic.  At mid-convection
zone the Reynolds number of the convection is $\sim 170$ and the
local Rossby number is $\sim 0.4$.

The differential rotation is established by Reynolds stresses in the
convection.  There is a monotonic decrease in local angular velocity
from the fast equator to the slow poles (Fig.~\ref{fig:vr_dr_and_T}$c$). 
Convective enthalpy transport in latitude establishes a prominent
latitudinal temperature contrast (Fig.~\ref{fig:vr_dr_and_T}$d$) as
convective cells partially align with the rotation axis.  This thermal
profile is consistent with a thermal wind balance serving to maintain
the differential rotation.

\begin{figure*}[t]
   \includegraphics{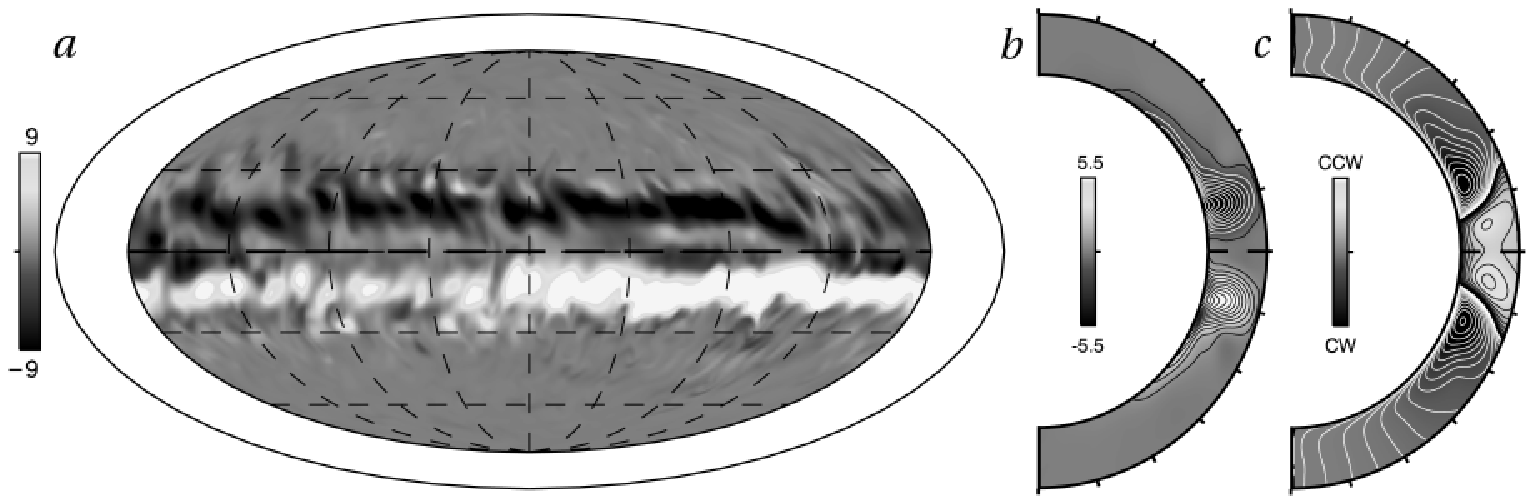}
\caption{Structure of magnetic field for case~D3: $(a)$ Snapshot of $B_\phi$ in
  Mollweide view 
  at mid-convection zone ($r=0.85 R_\sol$) with scale in kG,
  at the same instant as in Figure~\ref{fig:vr_dr_and_T}$a,b$.
  Strong magnetic domains have formed in the midst of the turbulent
  convection, possessing mainly opposite polarity in the two hemispheres. 
  $(b)$~Azimuthally averaged $B_\phi$ (scale in kG), with radius and latitude.  This
  has been further averaged over a period of 200 days.  The sea-snakes
  of magnetic field 
  evident in $(a)$ are present throughout the convection zone,
  with strongest average magnetic field amplitudes ($\sim 15\mathrm{kG}$) near the base of the
  convection zone. 
  $(c)$~Average poloidal magnetic field lines, showing the global
  structure of the mean field with polarity of field indicated (light
  tones counter-clockwise, dark tones clockwise).  
  The peak average amplitude of the
  poloidal field is about 6.5kG at the base of the convection zone. 
  \label{fig:magnetic_fields}}
\vspace{0.25cm}
\end{figure*}

\subsection{4. Strong magnetism amidst turbulent convection}
A remarkable finding in these rapidly rotating stellar dynamos is the
presence of strong, organized magnetic structures which fill the
convection zone and live amidst the turbulent convection.  The
structure of the toroidal field is illustrated in
Figure~\ref{fig:magnetic_fields}$a,b$.  At mid-convection zone, large
magnetic domains of each polarity appear near the equator.  The
magnetic fields in these ``sea-snakes'' are quite strong, with peak amplitudes over~20kG.
The mean toroidal fields are also quite strong, with typical strengths
of $\pm 5$kG and peak amplitudes of $\pm 15$kG.  This is in contrast
to previous solar simulations with ASH \citep{Brun_et_al_2004} where
the magnetic fields were dominated by fluctuating components and the mean
fields were quite weak.  In the solar case our simulations have
required a tachocline of penetration and shear
\citep{Browning_et_al_2006} to generate similarly ordered toroidal
structures.  Here we have no tachocline and the fields are generated
in the bulk of the convection zone.  The sea-snakes of case~D3 are
quite long lived, persisting for thousands of days despite the 
turbulent convection.  The average poloidal fields are
much weaker than the toroidal fields, with typical strengths of
$\pm3$kG and peak amplitudes of~$\pm6$kG.  The global field has
prominent dipolar and quadrapolar components, as shown in
Figure~\ref{fig:magnetic_fields}$c$.  The mean poloidal and toroidal
fields are relatively steady in time and this dynamo shows no cyclic
behaviour or oscillations in differential rotation and magnetism.

\begin{figure*}[t]
  \includegraphics[width=\linewidth]{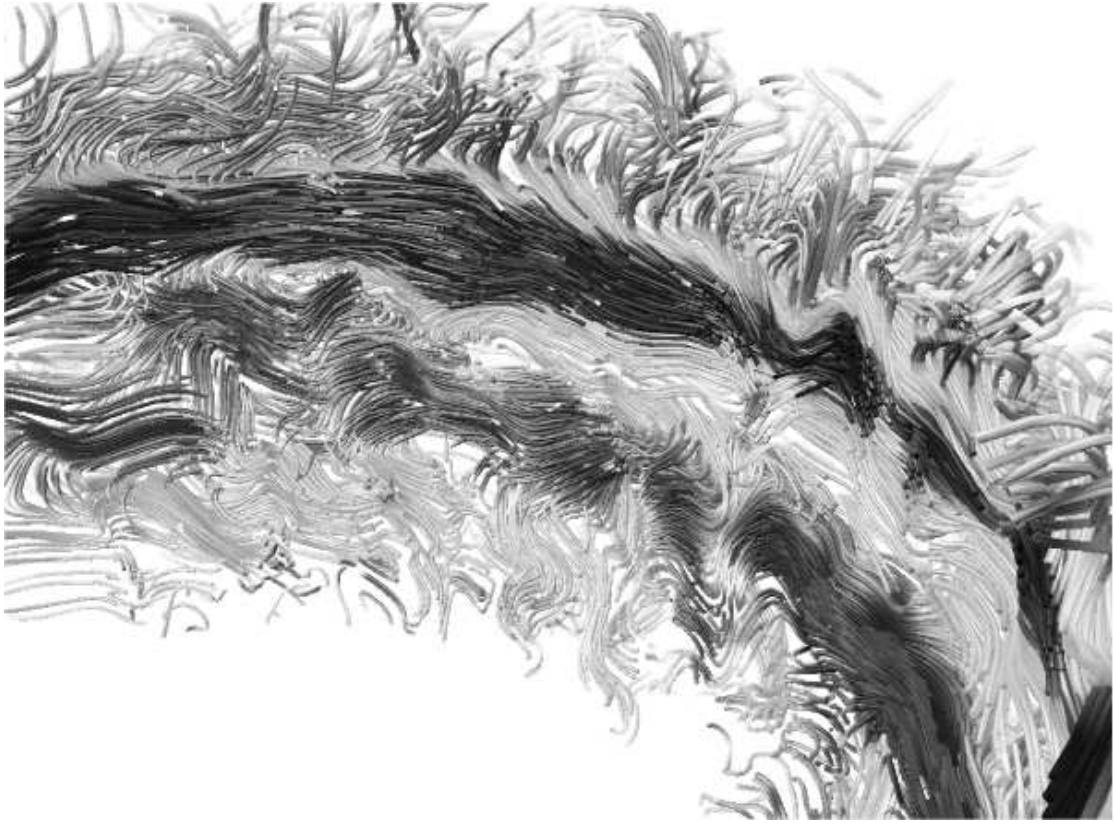}
  \caption{Volume rendering of magnetic fields filling the bulk of the
  convection zone in case~D3, showing that these flux concentrations are complex,
  with field lines threading in and out of the concentrated regions.
  The view is from the center of the star and directed towards the
  surface, and occupies a region near the equator that spans from $\pm
  45^\circ$ in latitude.  The two regions of strong opposite polarity (dark
  tones) above and below the equator are largely toroidal field, while
  the weaker fields reaching towards the polar regions have been
  twisted by the vortical convection.
    \label{fig:volume_rendering}}
\end{figure*}

The three-dimensional structure of the magnetic field is rich and
complex (as shown in Fig.~\ref{fig:volume_rendering}).  The sea-snakes are dominantly toroidal field, but fields
thread in and out of these structures and connect to the higher
latitudes.  Strong downflows kink the sea-snakes in many locations,
dragging field to the base of the convection zone before it rejoins
the larger structure.  Some magnetic fields span across the equatorial region,
connecting the two zones of opposite polarity.

After the dynamo has saturated, the total volume-averaged kinetic
energy density of motions relative to the rotating reference frame is
about $7.15\times 10^{6}~\mathrm{ergs}~\mathrm{cm}^{-3}$. 
Much of this is contained in the mean differential rotation which
comprises about 69\% of the total. Fluctuating convective motions
contain almost 31\% of the total and meridional circulations are a minor component.  
The total volume-averaged magnetic energy density is about
11\% of the kinetic energy density.   Of this, 48\% is contained in
the mean toroidal fields, 47\% in the fluctuating fields
and about 5\% in the poloidal fields.

\begin{figure*}[t]
  \includegraphics{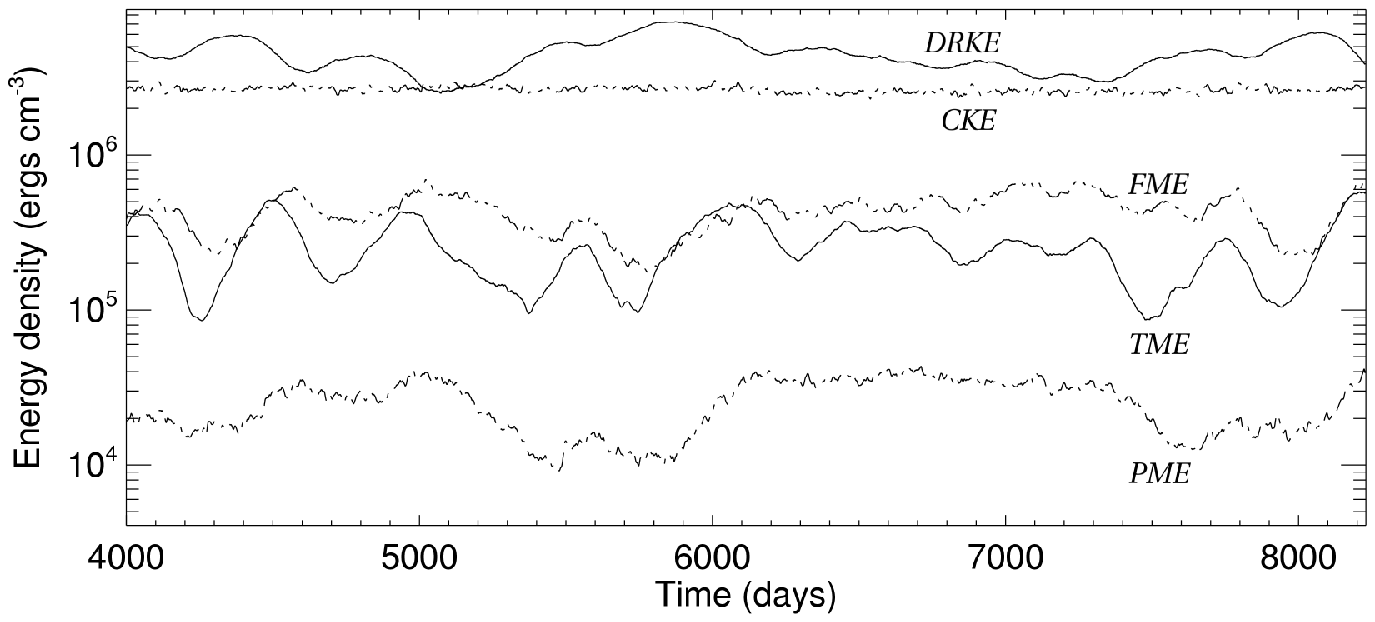}
  \caption{Volume averaged energy densities late in case~D3$a$.
  Shown are kinetic energies in differential rotation (DRKE) and
  convection (CKE) as well as magnetic energies in mean toroidal
  fields (TME), mean poloidal fields (PME) and in
  the fluctuating fields (FME).  Oscillations between kinetic energy
  and magnetic energies occur, with DRKE and PME moving oppositely
  over long periods and TME and FME moving largely together on shorter
  timescales.
\label{fig:energy_trace}}
\end{figure*}

\subsection{5. Cyclic behavior amidst stronger turbulence}
Our current models are far in parameter space from the levels of
turbulence realized in the convection zones of real stars.  A natural
question therefore is what occurs as we drive these turbulent
solutions to even higher levels of complexity.  Are stronger magnetic
fields generated?  Do the persistent magnetic structures found in the
equatorial regions survive the more vigorous turbulence?

To begin addressing these questions we have conducted a second dynamo
simulation for a similar star rotating three times the solar rate but with lower
levels of eddy diffusivities and therefore higher levels of
convective turbulence.  The diffusivities in our original dynamo
case~D3 were dropped by about 30\% to a new mid-convection zone value
of $\nu$ of $0.94 \times 10^{12}~\mathrm{cm}^2/\mathrm{s}$, while
maintaining a Prandtl number of 0.25 and a magnetic Prandtl number of
0.5.  The new solution, case~D3$a$, has a mid-convection Reynolds
number of $\sim 230$ and a local Rossby number of $\sim 0.45$.

An intriguing result from this simulation is that the dynamo has
achieved a new state, wherein oscillations in energy and magnetic field
structure occur over time scales which are long in comparison to
characteristic convective turn-over times or the stellar rotation
rate, as is illustrated in Figure~\ref{fig:energy_trace}. The kinetic
energy in differential rotation undergoes similar 
long-period oscillations, whereas the kinetic energy in convection and
the meridional circulations is nearly unaffected.  The structure of
the magnetic fields is similar to case~D3, with two strong sea-snakes
of opposite polarity above and below the equator.  With the
oscillations in magnetic energy, the sea-snakes wax and wane in
strength, sometimes synchronously and sometimes individually,
with one polarity growing and one fading.

The total kinetic energy relative to the rotating reference frame is
somewhat lower in this simulation, with a volume and time average of 
$6.82 \times 10^{6}~\mathrm{ergs}~\mathrm{cm}^{-3}$ over the 4200 day period
shown in Figure~\ref{fig:energy_trace}.  Mean differential rotation
accounts for 60\% of this, while convection accounts for nearly 40\%.
The time averaged total magnetic energy is about 11\% of the kinetic
energy, with mean toroidal fields accounting for about 33\% of the
total magnetic energy and fluctuating fields comprising nearly 63\%.
There are intervals where the sense of the large-scale toroidal fields
nearly flips, but then the fields regain strength while retaining
their original sense.

\subsection{6. Global dynamos without a tachocline}
A surprising result of these simulations of rapidly rotating stars is
that global-scale toroidal and poloidal magnetic fields can be built
and maintained in the bulk of the convection zone, despite the
presence of turbulent convective motions, without resorting to a
stable tachocline of shear which stores and combs out the large-scale
fields.  For the solar dynamo,
it has long been thought that the turbulence of the convection zone
precludes the existence of globally organized fields there.  Instead, we 
have arrived at the theoretical concept of an interface dynamo \citep{Parker_1993}
in which fluctuating fields arise from dynamo action in the convection
zone, and these are then
pumped into the tachocline where gradually global-scale fields are
built and organized.  Indeed, previous ASH simulations of solar dynamo
action in the bulk of the convection zone \citep{Brun_et_al_2004}
built mostly fluctuating fields with little global-scale structure.
Organized, global-scale toroidal fields have only emerged in
solar simulations where a tachocline has been included
\citep{Browning_et_al_2006}.  These simulations of rapidly rotating
solar-like stars represent the first time to our knowledge when
organized structures have been built and maintained amidst vigorous,
three-dimensional turbulent convection in the bulk of the convection
zone.  

These simulations suggest that dynamo action in rapidly rotating stars may yield
global-scale fields even without resorting to an interface dynamo.
This latter dynamo crucially requires a boundary layer of strong
radial shear to comb the fields into strong toroidal structures.  Yet
with more rapid rotation this seems to be no longer necessary.
This may have important implications for fully convective stars in which no such
interface exists.  The sea-snakes of toroidal field
arising in the current simulations are able to freely rise
through our upper boundary, yet they remain strong and fill the convection
zone.  The pumping action of convective downflows near the equator
must effectively resist the magnetic buoyancy which otherwise would
lead to these fields escaping the domain.
The behavior exhibited here is changing our intuition of the dynamo
action that can be realized in stars with rapid rotation.
A crucial question for the future is whether these
organized magnetic structures survive in the convection zone in the
presence of a penetrative tachocline

\vspace{0.5cm}

This research was supported by NASA through Heliophysics Theory
Program grant NNG05G124G and the NASA GSRP program by award number
NNG05GN08H.  The simulations were carried out with NSF PACI support of
PSC and SDSC.

\bibliographystyle{mn2e}   
\def\aap{A\&A}
\def\apj{ApJ}
\def\apjl{ApJ}
\def\araa{ARA\&A}
\bibliography{bibliography}

\begin{thebibliography}{}

\bibitem[\protect\citeauthoryear{{Brown}, {Browning}, {Brun}, {Miesch} \&
  {Toomre}}{{Brown} et~al.}{2007}]{Brown_et_al_2007}
{Brown} B.~P.,  {Browning} M.~K.,  {Brun} A.~S.,  {Miesch} M.~S.,    {Toomre}
  J.,  2007, \apj, submitted

\bibitem[\protect\citeauthoryear{{Browning}, {Miesch}, {Brun} \&
  {Toomre}}{{Browning} et~al.}{2006}]{Browning_et_al_2006}
{Browning} M.~K.,  {Miesch} M.~S.,  {Brun} A.~S.,    {Toomre} J.,  2006, \apjl,
  648, L157

\bibitem[\protect\citeauthoryear{{Brun}, {Antia}, {Chitre} \& {Zahn}}{{Brun}
  et~al.}{2002}]{Brun_et_al_2002}
{Brun} A.~S.,  {Antia} H.~M.,  {Chitre} S.~M.,    {Zahn} J.-P.,  2002, \aap,
  391, 725

\bibitem[\protect\citeauthoryear{{Brun}, {Miesch} \& {Toomre}}{{Brun}
  et~al.}{2004}]{Brun_et_al_2004}
{Brun} A.~S.,  {Miesch} M.~S.,    {Toomre} J.,  2004, \apj, 614, 1073

\bibitem[\protect\citeauthoryear{{Charbonneau}}{{Charbonneau}}{2005}]{Charbonn%
eau_2005}
{Charbonneau} P.,  2005, Living Reviews in Solar Physics, 2, 2

\bibitem[\protect\citeauthoryear{{Clune}, {Elliott}, {Glatzmaier}, {Miesch} \&
  {Toomre}}{{Clune} et~al.}{1999}]{Clune_et_al_1999}
{Clune} T.~L.,  {Elliott} J.~R.,  {Glatzmaier} G.~A.,  {Miesch} M.~S.,
  {Toomre} J.,  1999, Parallel Computing, 25, 361

\bibitem[\protect\citeauthoryear{{Donahue}, {Saar} \& {Baliunas}}{{Donahue}
  et~al.}{1996}]{Donahue_et_al_1996}
{Donahue} R.~A.,  {Saar} S.~H.,    {Baliunas} S.~L.,  1996, \apj, 466, 384

\bibitem[\protect\citeauthoryear{{Parker}}{{Parker}}{1993}]{Parker_1993}
{Parker} E.~N.,  1993, \apj, 408, 707

\bibitem[\protect\citeauthoryear{{Thompson}, {Christensen-Dalsgaard}, {Miesch}
  \& {Toomre}}{{Thompson} et~al.}{2003}]{Thompson_et_al_2003}
{Thompson} M.~J.,  {Christensen-Dalsgaard} J.,  {Miesch} M.~S.,    {Toomre} J.,
   2003, \araa, 41, 599

\end{thebibliography}

\end{document}